\documentclass[10pt,draftcls,onecolumn]{IEEEtran}
\usepackage{graphicx}
\usepackage{amssymb,amsmath}
\usepackage{amsfonts}
\usepackage{epstopdf}
\usepackage{mathrsfs}
\usepackage{array}
\graphicspath{{figs/}}
\DeclareMathAlphabet{\mathpzc}{OT1}{pzc}{m}{it}
\usepackage{cite}
\usepackage[ruled,vlined]{algorithm2e}
\usepackage[para,online,flushleft]{threeparttable}

\usepackage{algpseudocode}

\begin{document}
\title{Enhanced Factored Three-Way Restricted Boltzmann Machines for Speech Detection}
\author{Pengfei~Sun,~\IEEEmembership{Student Member,}
        and~Jun~Qin,~\IEEEmembership{Member,~IEEE,}} 
\markboth{Under consideration at Pattern Recognition Letters, April~2017}%
{Under consideration at Pattern Recognition Letters, April~2017}
\maketitle
\begin{abstract} In this letter, we propose enhanced factored three-way restricted Boltzmann machines (EFTW-RBMs) for speech detection. The proposed model incorporates conditional feature learning by introducing a multiplicative input branch, which allows a modulation over visible-hidden node pairs. Instead of directly feeding previous frames of speech spectrum into this third unit, a specific algorithm, including weighting coefficients and threshold shrinkage, is applied to obtain more correlated and representative inputs. To reduce the parameters of the three-way network, low-rank factorization is utilized to decompose the interaction tensor, on which non-negative constraint is also imposed to address the sparsity characteristic. This enhanced model effectively strengthens the structured feature learning that helps to recover the speech components from noisy background. The validations based on the area-under-ROC-curve (AUC) and signal distortion ratio (SDR) show that our EFTW-RBMs outperforms several existing 1D and 2D (i.e., time and time-frequency domain) speech detection algorithms in various noisy environments. 
\end{abstract}
\IEEEpeerreviewmaketitle

\section{Introduction}
\IEEEPARstart{S}{peech} detection (SD) greatly improves the separation of speech sources from background interferes \cite{sohn1999statistical}. Nowadays, SD techniques attract intense attentions in a general speech processing framework, including automatic speech recognition (ASR) \cite{li2014overview}, speech enhancement \cite{gerkmann2012unbiased} and speech coding \cite{sohn1999statistical}. 

Recently, deep neural network (DNN) based 1D SD algorithms show great advantages over conventional voice activity detectors \cite{zhang2013deep, sadjadi2013unsupervised}. The obvious benefits of such approaches lie on their easy integration into ASR, robust performance, and feature fusion capability. Zhang and Wu \cite{zhang2013deep} introduced deep belief network and used stacked Bernoulli-Bernoulli restricted Boltzmann machines (RBMs) to conduct the 1D SD. The idea that incorporating temporal context correlation to strengthen the dynamical detection is widely used in network structure design \cite{leglaive2015singing, eyben2013real}. Other DNN based 1D SD strategies might either focus on improving the front-end acoustic feature inputs (e.g., acoustic models and statistical models) \cite{ryant2013speech, hwang2016ensemble}, or exploiting the supervised network structure in terms of sample training \cite{thomas2015improvements}. These DNN based approaches rely on comprehensive network training, and then are applied to binarily label the speech activities in the time domain. However, 1D SD methods integrate frequency features, and cannot reveal information in the joint time-frequency domain, which are generally more expressive on speech activities, compared with the binary values in 1D SD approaches. 

In this study, we propose enhanced factored three-way RBMs (EFTW-RBMs) for both 1D and 2D SD. The main idea is utilizing backward sampling from the hidden layer that represents the speech features. The well-trained RBMs can effectively reconstruct the speech components, which further are normalized as the speech presence probability (SPP) values. The proposed EFTW-RBMs model introduces a multiplicative third branch to exploit the strong correlations in consecutive speech frames, therefore can more effectively to capture the speech structures. A continuously updated memorized input is provided by applying weighting coefficients $\alpha$ and threshold function $\mathpzc{S}_{T}$, which retains global frames based on the locally updated three-way RBMs. Low rank factorization and non-negative regularization are applied for the network training. 

\section{Proposed Method}
\subsection{Gated Restricted Boltzmann Machines} 
In previous study \cite{memisevic2010learning}, multiplicative gated RBMs are described by an energy function that captures correlations among the components of $x$, $y$ and $h$ 
\begin{equation}
\begin{split}
E(\mathbf{y},\mathbf{h};\mathbf{x}) = &-\sum_{ijk}w_{ijk}\frac{x_{i}}{\sigma_{i}}\frac{y_{j}}{\sigma_{j}}h_{k}-\sum_{k}w_{k}^{h}h_{k}\\
&+\sum_{j}\frac{(y_{j}-w_{j}^{y})^{2}}{2\sigma_{j}^{2}}
\end{split} 
\label{energy}
\end{equation}
where $i$, $j$ and $k$ index input, visible and hidden units, respectively. The bold font represents the variable, and small cap denotes the observation. $x_{i}$ and $y_{j}$ are Gaussian units, and $h_{k}$ is the binary state of the hidden unit $k$. $\sigma_{i}$ and $\sigma_{j}$ are the standard deviations associated with $x_{i}$ and $y_{j}$, respectively. The components $w_{ijk}$ of a three-way tensor connect units $x_{i}$, $y_{j}$ and $h_{k}$. The terms $w_{k}^{h}$ and $w_{j}^{y}$ represent biases of the hidden and visible units, respectively. The energy function assigns a probability to the joint configuration as: 
\begin{equation}
p(\mathbf{y},\mathbf{h}|\mathbf{x}) = \frac{1}{Z(\mathbf{x})}exp(-E(\mathbf{y},\mathbf{h};\mathbf{x}))
\end{equation}
\begin{equation}
Z(\mathbf{x}) = \sum_{\mathbf{h},\mathbf{y}}exp(-E(\mathbf{y},\mathbf{h};\mathbf{x}))
\end{equation}
where the normalization term $Z(\mathbf{x})$ is summed over $\mathbf{y}$ and $\mathbf{h}$, and hence defining the conditional distribution $p(\mathbf{y},\mathbf{h}|\mathbf{x})$. Since there is no connections between the neurons in the same layer, inferences of the $k$th hidden and $j$th visible unit can be performed as
\begin{equation}
p(h_{k}=1|\mathbf{y};\mathbf{x}) = S(\Delta E_{k})
\label{eqhp}
\end{equation}
\begin{equation}
p(y_{j} = y|\mathbf{h};\mathbf{x}) = N(y|\Delta E_{j},\sigma_{j}^{2})
\label{eqvp}
\end{equation}
where $N(\cdot|\mu, \sigma^{2})$ denotes the Gaussian probability density function with mean $\mu$ and standard deviation $\sigma$. $S(\cdot)$ is the sigmoid activation function. $\Delta E_{k}$ and $\Delta E_{j}$ are the overall inputs of the $k$th hidden unit and $j$th visible unit, respectively \cite{yamashita2014bernoulli}. The gated RBMs allows the hidden units to model the transition between successive frames.  

\subsection{Factored Three-Way Restricted Boltzmann Machines} 
\begin{figure}[!hbt]
\vspace{-4mm}
\centerline{\includegraphics[scale=0.65]{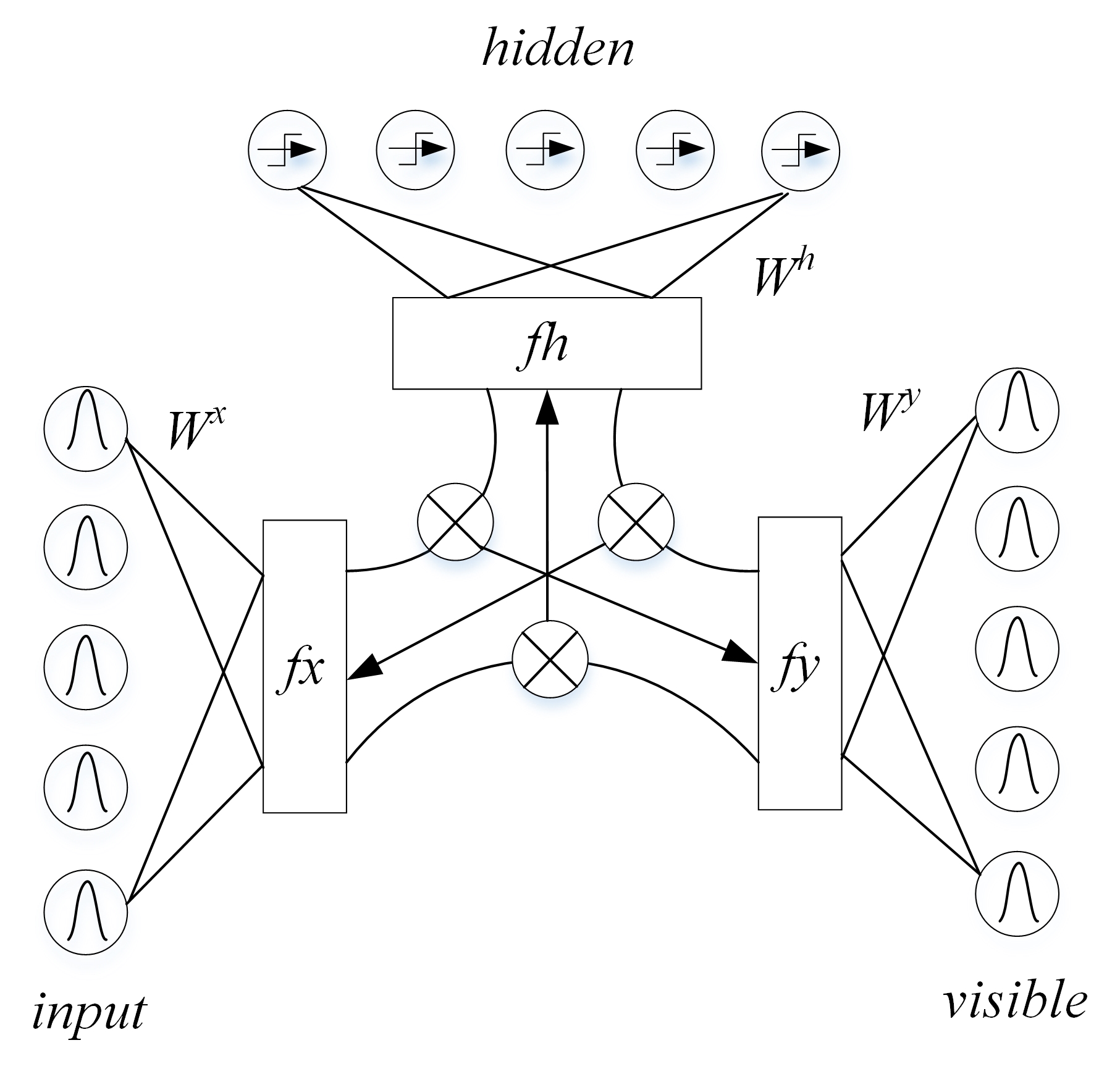}}
\vspace{-3mm}
\caption{The schematic of symmetrical three-way RBMs. The three factor layers have the same size, and $\otimes$ refers to element-wise multiplication.}
\vspace{-3mm}
\label{recursiveNN}
\end{figure}

To reduce the parameters of three-way RBMs, the interaction tensor $w_{ijk}$ can be factored into decoupled matrices \cite{memisevic2010learning}. In our study, in order to reconstruct the inputs to obtain the weighting coefficients $\alpha$, a symmetrical structure is proposed as shown in Fig.~\ref{recursiveNN}. As a result, the bias term $w_{i}^{x}$ is added to the input unit $x_{i}$ to balance the structure of visible unit $y_{j}$, and FTW-RBMs in the energy function ~(\ref{energy}) can be rewrite as
\begin{equation}
\begin{split}
-E(\mathbf{y},\mathbf{h};& \mathbf{x}) = -\sum_{i}\frac{(x_{i}-w_{i}^{x})^{2}}{2\sigma_{i}^{2}} -\sum_{j}\frac{(y_{j}-w_{j}^{y})^{2}}{2\sigma_{j}^{2}}\\
 & \sum_{k}w_{k}^{h}h_{k}+ \sum_{f=1}^{F}\sum_{ijk}w_{if}^{x}w_{jf}^{y}w_{kf}^{h}\frac{x_{i}}{\sigma_{i}}\frac{y_{j}}{\sigma_{j}}h_{k} 
\end{split}
\label{eqfactor}
\end{equation}
where the $I\times J \times K$ parameter tensor $w_{ijk}$ is replaced by three matrices (i.e., $w_{if}^{x}$, $w_{jf}^{y}$, and $w_{kf}^{h}$) with sizes $I\times F$, $J\times F$ and $K\times F$, in which $f$ is the factor index. Accordingly, it can be reorganized into 
\begin{equation}
\begin{split}
&-E(\mathbf{y},\mathbf{h};\mathbf{x}) = \sum_{k}w_{k}^{h}h_{k}-\sum_{j}\frac{(y_{j}-w_{j}^{y})^{2}}{2\sigma_{j}^{2}}- \sum_{i}\frac{(x_{i}-w_{i}^{x})^{2}}{2\sigma_{i}^{2}}\\
&+ \sum_{f}\left(w_{if}^{x}\sum_{i}\frac{x_{i}}{\sigma_{i}}\right) \left(\sum_{j} w_{jf}^{y}\frac{y_{j}}{\sigma_{j}} \right) \left( \sum_{k}w_{kf}^{h}h_{k}\right)
\end{split}
\end{equation}
By noting $f_{f}^{x}= \sum_{i=1}^{I}w_{if}^{x}\frac{x_{i}}{\sigma_{i}}, \quad  f_{f}^{y}= \sum_{j=1}^{J}w_{jf}^{y}\frac{y_{j}}{\sigma_{j}}, \quad f_{f}^{h}= \sum_{k=1}^{K}w_{kf}^{h}h_{k}$. The three factor layers as shown in Fig.~\ref{recursiveNN} have the same size $F$, and the factor terms (i.e., $W^{x}\mathbf{x}$, $W^{y}\mathbf{y}$, and $W^{h}\mathbf{h}$) correspond to three linear filters applied to the input, visible, and the hidden unit, respectively. To perform k-step Gibbs sampling in the factored model, the overall inputs of each unit in the three layers are calculated as
\begin{equation}
\Delta E_{k} = \sum_{f}w_{kf}^{h}\sum_{i}w_{if}^{x}\frac{x_{i}}{\sigma_{i}}\sum_{j}w_{jf}^{y}\frac{y_{j}}{\sigma_{j}}+ w_{k}^{h}
\label{hidden}
\end{equation} 
\begin{equation}
\Delta E_{j} = \sum_{f}w_{jf}^{y}\sum_{i}w_{if}^{x}\frac{x_{i}}{\sigma_{i}}\sum_{k}w_{kf}^{h}h_{k}+ w_{j}^{y}
\label{visible}
\end{equation}
\begin{equation}
\Delta E_{i} = \sum_{f}w_{if}^{x}\sum_{j}w_{jf}^{y}\frac{y_{j}}{\sigma_{j}}\sum_{k}w_{kf}^{h}h_{k}+w_{i}^{x}
\label{input}
\end{equation}

In (\ref{hidden})-(\ref{input}), the factor layers are multiplied element-wise (as the $\otimes$ illustrated in Fig.~\ref{recursiveNN}) through the same index $f$. These are then substituted in (\ref{eqhp})-(\ref{eqvp}) for determining the probability distributions for each of the visible and hidden units. For input units, the symmetrical form $N(x|\Delta E_{i},\sigma_{i}^{2})$ is used. The FTW-RBMs model learn speech patterns in the hidden units by pairwise matching input filter responses and visible filter responses, and this procedure aims to find a set of filters that can reflect the correlations of consecutive speech frames in the training data.

\subsection{Enhanced Input Units} 
\begin{figure}[!hbt]
\vspace{-4mm}
\centerline{\includegraphics[scale=1.3]{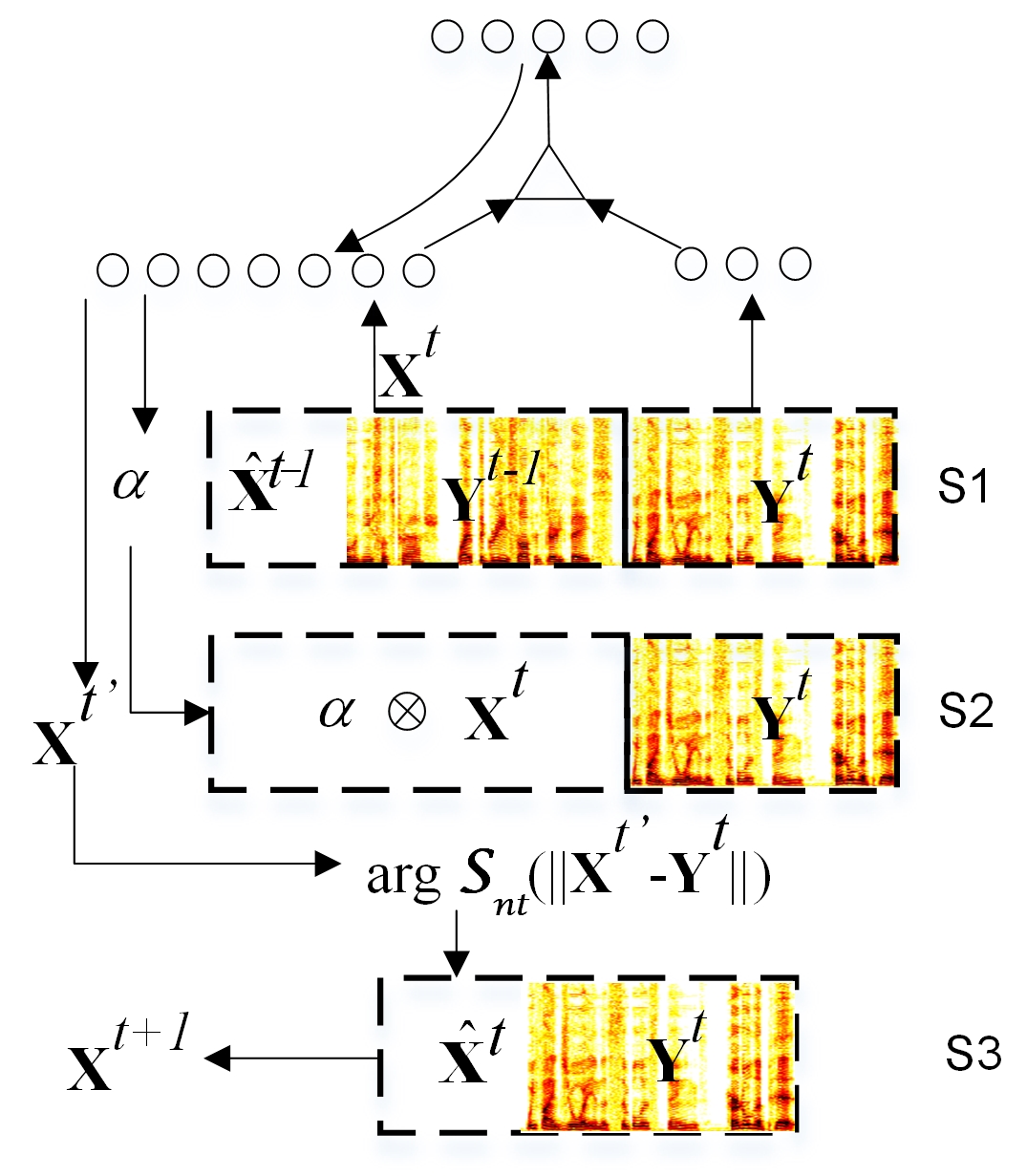}}
\vspace{-3mm}
\caption{The schematic of EFTW-RBMs. In a batch training loop, it includes three steps, referring as S1, S2, S3. The arrow indicates the data flow direction. In the network, arrow with up direction means forward proper}
\vspace{-3mm}
\label{Recursive}
\end{figure}

In FTW-RBMs model described in section B, the consecutive frames of noisy speech are fed into the trained networks as input and visible units, respectively. The SD results reflected by SPP distribution are obtained from the reconstructed visible units. The multiplicative structure of FTW-RBMs helps to amplify the speech features, however, unavoidably introduces high sensitivity to transient noise components. To obtain a more robust network, our EFTW-RBMs model selectively retains long-term speech features by proposing an enhanced input as shown in Fig.~\ref{Recursive}, in which three steps (i.e., S1, S2, and S3) are implemented in each data batch training loop. 

In S1, $\mathbf{X}^{t} \in \mathcal{R}^{n\times (m+n_{t})}$ and $\mathbf{Y}^{t} \in \mathcal{R}^{n\times m}$ are fed into the input and visible units ($t$ index different training batches in the time domain), where $n_{t}$ is the globally retained frames. $\mathbf{X}^{t} = [\hat{\mathbf{X}}^{t-1} \mathbf{Y}^{t-1}]$, in which $\mathbf{Y}^{t-1}$ are the visible units in $(t-1)$th training loop, and $\hat{\mathbf{X}}^{t-1} \in \mathcal{R}^{n\times n_{t}}$ is selected from previous input units $\mathbf{X}^{t-1}$. The backward sampling in terms of the input branch generates the weight coefficients $\alpha$, which can be given as
\begin{equation}
\alpha = N(\mathbf{x}^{t}|\Delta E_{i}, \sigma_{i}^{2}) 
\label{weighting}
\end{equation} 
(\ref{weighting}) represents the normalized reconstruction of the input data batch, in which both hidden layer and visible layer are involved. As a result, $\alpha$ reflects the correlation between the consecutive training units, and is also subjected to the inverse constraint imposed by the visible unit $\mathbf{y}^{t}$. 

In S2, element-wise multiplied $\alpha\mathbf{X}^{t}$ is used as the input for network training. The reconstructed input units are notated as $\mathbf{X}^{t'}$. Accordingly, we define $\lambda_{i} = \min_{j\leq m}||\mathbf{X}^{t'}(:,i)- \mathbf{Y}^{t}(:,j)||$, $i \leq (n_{t} +m)$. $\lambda_{i}$ indicates the distance between $i$th column vector of $\mathbf{X}^{t}$ and the matrix $\mathbf{Y}^{t}$. The smaller value $\lambda_{i}$ is, the closer to $\mathbf{Y}^{t}$ the vector is. To retain the global speech features, a criterion is proposed to obtain $\hat{\mathbf{X}}^{t}$ by selecting column vectors from $\mathbf{X}^{t}$, that is described as
\begin{equation}
\hat{\mathbf{X}}^{t} = \{\mathbf{X}^{t}(:,i)|i \in \mathpzc{S}_{n_{t}}(\mathpzc{D}) \}
\label{threshold}
\end{equation} 
where $\mathpzc{S}_{n_{t}}(\cdot)$ is a threshold function that returns the index of $n_{t}$ smallest elements in $\mathpzc{D} = \{\lambda_{i}\}$. In S3 step, the input units $\mathbf{X}^{t+1} = [\hat{\mathbf{X}}^{t}  \mathbf{Y}^{t}]$ at $(t+1)$th data batch are prepared for the next training loop.

The enhanced input units in the proposed EFTW-RBMs include a dynamical bias through the weighting coefficients $\alpha$, whereas the backward sampling approach imposes the constraints of visible units. The threshold function also provides a benefit to retain the global features of speech, similar to the long short-term memory method \cite{leglaive2015singing}. When implementing the SD, the noisy speech are segmented into consecutive data batches as input and visible units, and accordingly the proposed EFTW-RBMs model can backwardly generate the reconstructed visible layer, which can be normalized as SPP.

\subsection{Probabilistic inference and learning rules} 
To train EFTW-RBMs, one needs to maximize the average log-probability $L=$log $p(\mathbf{y}|\mathbf{x})$ of a set of training pairs $\{(\mathbf{x}, \mathbf{y})\}$. The derivative of the negative log-probability with respect to parameters $\theta$ is given as 
\begin{equation}
-\frac{\partial L}{\partial \theta} = \langle \frac{\partial E(\mathbf{y}, \mathbf{h}; \mathbf{x})}{\partial \theta} \rangle_{\mathbf{h}}-\langle \frac{\partial E(\mathbf{y}, \mathbf{h}; \mathbf{x})}{\partial \theta} \rangle_{\mathbf{h},\mathbf{y}}
\label{eqderi}
\end{equation}
where $\langle \rangle_{\mathbf{v}}$ denotes the expectation with respect to variable $\mathbf{v}$. In practical, Markov chain step running is used to approximate the averages in Eq.~(\ref{eqderi}). By differentiating (\ref{eqfactor}) with respect to the parameters, we get 
\begin{equation}
-\frac{\partial E}{\partial w_{kf}^{h}} = -h_{k}\sum_{i}x_{i} w_{if}^{x}\sum_{j}y_{j}w_{jf}^{y}
\label{liklyhood1}
\end{equation}
\begin{equation}
-\frac{\partial E}{\partial w_{jf}^{y}} = -y_{j}\sum_{i}x_{i}w_{if}^{x}\sum_{k}h_{k}w_{kf}^{h}
\label{liklyhood2}
\end{equation}
\begin{equation}
-\frac{\partial E}{\partial w_{if}^{x}} = -x_{i}\sum_{j}y_{j}w_{jf}^{y}\sum_{k}h_{k}w_{kf}^{h}
\label{liklyhood3}
\end{equation}
\begin{equation}
-\frac{\partial E}{\partial w_{k}^{h}} = h_{k}, \quad -\frac{\partial E}{\partial w_{i}^{x}} = x_{i}, \quad -\frac{\partial E}{\partial w_{j}^{y}} = y_{j}
\label{liklyhood4}
\end{equation}

To encourage nonnegativity in three factor matrices $w_{kf}$, $w_{if}$, and $w_{jf}$, a quadratic barrier function is incorporated to modify the log probability. As a result, the objective function is the regularized likelihood illustrated as \cite{nguyen2013learning}
\begin{equation}
- \mathcal{L}_{reg}= \mathcal{L}(\mathbf{y};\mathbf{x})- \frac{\beta}{2}\sum_{}\sum_{}f(w)
\label{sparsity}
\end{equation}
where 
\[
    f(x)=\left\{
                \begin{array}{ll}
                  x^{2} \qquad x<0\\
                  0 \qquad x\geq 0
                \end{array}
              \right.
\]

Based on (18), the parameter update can be translated as 
\begin{equation}
w \leftarrow w+\eta(\langle \rangle_{h}-\langle \rangle_{h,y}-\beta \lceil w \rceil^{-})
\label{sparsity1}
\end{equation}
where $\lceil w \rceil^{-}$ denotes the negative part of the weight. The number of hidden units and factors should be in a region that exact number of hidden units and factors did not have a strong influence on the result. The procedure of EFTW-RBMs for SD can be illustrated as following: 
\begin{algorithm} 
  \caption{EFTW-RBMs for SD} 
  \textbf{Training} \\
 \For{iteration $\leq$ $N_{epoch}$}   
  {
    \For{Iteration $\leq$ $N_{batch}$}
    {
        $y^{t}$ = $\mathbf{Y}^{t}$\; 
        $x^{t}$ = [$\hat{\mathbf{X}}^{t-1}$ $\mathbf{Y}^{t}$]\;
        sample \(h^{t} \sim p(h|y^{t},x^{t})\) by (\ref{eqhp})(\ref{hidden}) \;
        calculate $\langle\frac{\partial E}{\partial \theta}\rangle_{h}$ by (\ref{liklyhood1})-(\ref{liklyhood4})\; 
          \For {iteration $\leq$ $N_{step}$}
          {
              sample \(h^{t,n} \sim p(h|y^{t,n}, x^{t,n})\)  by (\ref{eqhp})(\ref{hidden})\; 
              sample \(y^{t,n} \sim p(y|x^{t,n}, h^{t,n})\)  by (\ref{eqvp})(\ref{visible})\;                
          } 
        calculate $\langle\frac{\partial E}{\partial \theta}\rangle_{h,y}$ by (\ref{liklyhood1})-(\ref{liklyhood4}) \; 
        update parameter set \(\{w_{kf}^{h}, w_{jf}^{y}, w_{if}^{x} ,w_{k}^{h},w_{j}^{y}, w_{i}^{x}\}\) by (\ref{eqderi}) and (\ref{sparsity1}) \; 
        sample \(\alpha^{t} \sim p(x|y^{t,n}, h^{t,n})\) by (\ref{input})\;
        update $\hat{\mathbf{X}}^{t}$ by (\ref{threshold})\;       
    }
  } 
  \textbf{SPP estimation} \\  
   \For{iteration $\leq$ $N_{frames}$} 
    { 
      sample \(h^{t} \sim p(h|y^{t}) \) by (\ref{hidden}) \;
      sample \(\alpha^{t} \sim p(x|y^{t}, h^{t})\) by (\ref{input})\;
      update $x^{t}$ by (\ref{threshold})\;
      sample $y^{t} \sim p(y|h^{t},x^{t})$ by (\ref{visible})\;
      $\mathcal{P}(:,t)$ =  $y^{t}$ \;
    }
  \KwOut{SPP matrix $\mathcal{P}$} 
\end{algorithm}
 
\section{Experimental evaluation}  
In our evaluation experiments, the clean speech corpus, consisting of 600 training sentences and 120 test sentences, is obtained from IEEE wide band speech database \cite{loizou2013speech}. Three typical noise samples (i.e., babble, Gaussian, and pink) from the NOIZUS-92 are used to synthesize noisy speech at three input signal-to-noise ratios (SNRs) (i.e., -5, 0, and 5 dB). All signals are resampled to 16 kHz sampling rate, and the spectrograms are calculated with a window length of 32 ms, and a hop of 16 ms.

Both of EFTW-RBMs and FTW-RBMs are used to calculate 2D SPP, and the corresponding 1D SD is obtained by integrating the 2D SPP values along the frequency axis. Three state-of-the-arts (i.e., MLP-DNN \cite{van2013robust}, Ying \cite{ying2011voice}, and DBN \cite{zhang2013deep}) are used in 1D SD evaluation. For 2D SPP evaluation, due to the lack of related DNN based algorithm, we compare EFTW-RBMs with two conventional 2D SPP estimators, including Gerkmann \cite{gerkmann2008improved}, and EM-ML \cite{sun2016low}. The basic network parameters of EFTW-RBMs and FTW-RBMs are set as: the number of visible units is 257, hidden units is 30, the number of epoches is 40, the learning rate is 0.001, the three-way factor $f$ is 60, the gated frame number $n_{t}$ is 6, the nonnegative coefficient $\beta$ is 0.5. and the momentum value is set as 0.1, and set to 0 after 20 epochs. The training and testing data are normalized. Area-under-ROC-curve (AUC) \cite{zhang2013deep} is applied for the evaluation of 1D SD, while speech distortion ratio (SDR) that used to evaluate 2D SPP estimation is defined as  
\begin{equation}
SDR = \frac{\sum\sum (Y(n,m)\mathcal{P}(n,m)-S(n,m))^{2}}{\sum\sum S(n,m)^{2}}
\end{equation}
where $S$ is the clean speech, $Y$ is the noisy speech, and $\mathcal{P}$ is the SPP.  
\begin{figure}[!hbt]
\vspace{-5mm}
\centerline{\includegraphics[scale=0.0595]{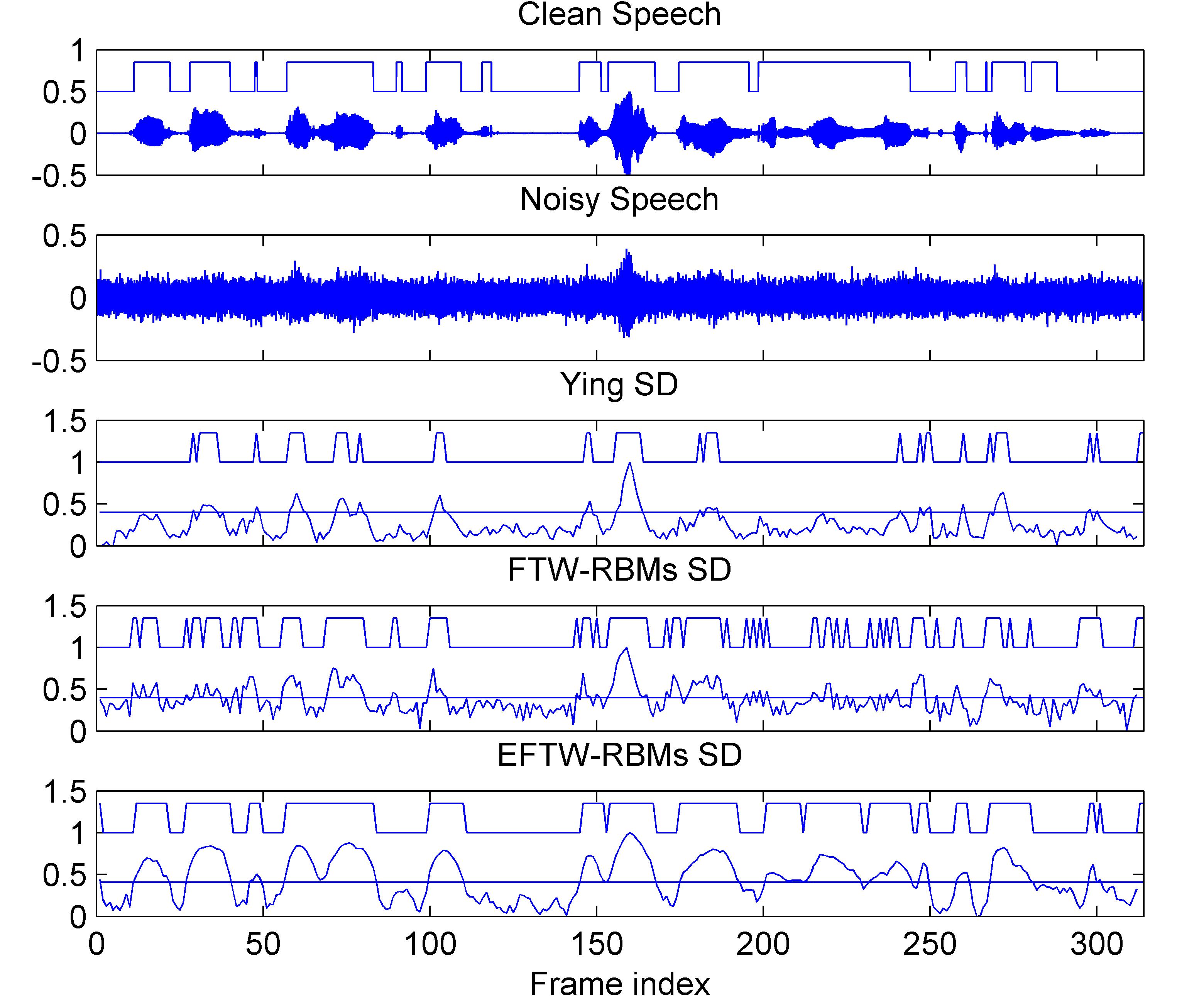}}
\vspace{-3mm}
\caption{Illustration of 1D SD by Ying, FTW-RBMs, and the proposed algorithm in pink noise environment at SNR=-5 dB. The output has been normalized to the range [0,1]. The straight lines are the optimal decision thresholds in terms of hit rate, and the notched lines show the hard decisions.}
\vspace{-2mm}
\label{1Ddetection}
\end{figure}

Figure \ref{1Ddetection} shows the intuitive evaluation of the 1D SD by Ying's SD model \cite{ying2011voice}, FTW-RBMs, and our EFTW-RBMs model in pink noise background at SNR = -5dB. The notched lines clearly demonstrate that EFTW-RBMs label the speech frames more accurately. Table~\ref{snoise1} summarizes the averaged AUC values obtained by five 1D SD algorithms in three noise at various SNRs. The performance of EFTW-RBMs is slightly higher than that of MLP-DNN algorithm, and obviously higher than that of Ying, FTW-RBMs, and DBN algorithms. 
\begin{center}
\begin{table}[ht]
\vspace{-5mm}
\caption{The AUC results for 1D speech detection. The results are averaged across all the speech utterances.} 
\centering 
\begin{threeparttable}
\scalebox{0.88}{
\begin{tabular*}{0.55\textwidth}{ c  c  c  c  c  c  c  c  c  c } 
\hline 
&\multicolumn{3}{c}{\footnotesize Babble} &\multicolumn{3}{c}{\footnotesize Gaussian} &\multicolumn{3}{c}{\footnotesize Pink} \\ 
&\footnotesize -5dB &\footnotesize 0dB &\footnotesize 5dB &\footnotesize -5dB &\footnotesize 0dB &\footnotesize 5dB &\footnotesize -5dB &\footnotesize 0dB &\footnotesize 5dB \\
\hline 
MLP-DNN        &0.79  &0.83  &0.86   &0.80  &0.85   &0.87  &0.78  &0.83  &0.86   \\
Ying           &0.61  &0.65  &0.69   &0.59  &0.62   &0.65  &0.62  &0.67  &0.70   \\
DBN            &0.75  &0.79  &0.82   &0.77  &0.61   &0.85  &0.72  &0.76  &0.81   \\
FTW-RBMs       &0.63  &0.71  &0.75   &0.79  &0.83   &0.85  &0.70  &0.74  &0.79   \\
\scriptsize EFTW-RBMs  &\textbf{0.80} &\textbf{0.84}  &\textbf{0.88} &\textbf{0.82}  &\textbf{0.86} &\textbf{0.89} &\textbf{0.81} &\textbf{0.85} &\textbf{0.88} \\ 
\hline 
\end{tabular*}
}
\end{threeparttable}
\vspace{-7mm}
\label{snoise1} 
\end{table}
\end{center}

\begin{figure}[!hbt]
\vspace{-4mm}
\centerline{\includegraphics[scale=0.067]{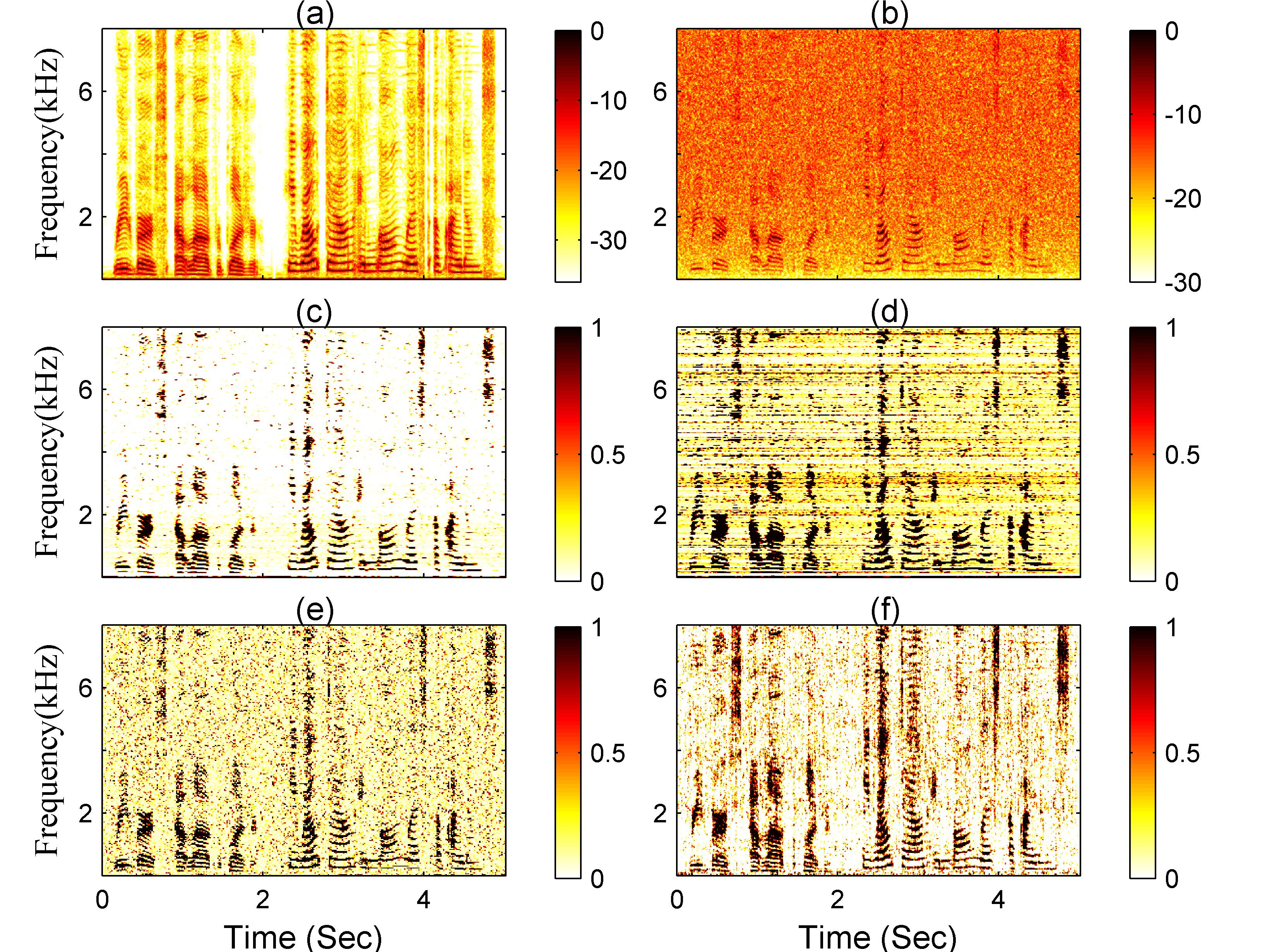}}
\vspace{-2mm}
\caption{The spectrograms of (a) clean speech, (b) noisy speech in pink noise at SNR= 0dB, and the 2D SPP results obtained by (c)EM-ML, (d)FTW-RBMs,(e)Gerkmann, and (f) the proposed EFTW-RBMs}
\vspace{-5mm}
\label{2Ddetection}
\end{figure}

Figure~\ref{2Ddetection} presents the 2D SPP results obtained by four 2D SD algorithms in pink noise at SNR = 0 dB. Unlike 1D SD that is labeled by binary values, 2D SD is presented as SPP ranging at [0,1]. Both EM-ML and Gerkmann models are based on statistical estimators, and the prior knowledge about noise should be obtained by assuming the initial frames as noise. The results show that our EFTW-RBMs successfully capture most of the speech activities, interpreted by the SPP values, which are proportional to the magnitudes of speech components. Moreover, Table \ref{snoise2} summarizes average SDR results of four 2D SD algorithms in three different noise at various SNRs. The bold numbers show that EFTW-RBMs obtain the lowest SDR in all three noise at different SNRs. It indicates that our proposed approach can successfully detect speech with less distortion compared with three other 2D SD algorithms.

\begin{center}
\begin{table}[ht]
\vspace{-3mm}
\caption{The SDR results for 1D and 2D speech detection. The results are averaged across all the speech utterances.} 
\centering 
\begin{threeparttable}
\scalebox{0.88}{
\begin{tabular*}{0.56\textwidth}{ c  c  c  c  c  c  c  c  c  c } 
\hline 
&\multicolumn{3}{c}{\footnotesize Babble} &\multicolumn{3}{c}{\footnotesize Gaussian} &\multicolumn{3}{c}{\footnotesize Pink} \\ 
&\footnotesize -5dB &\footnotesize 0dB &\footnotesize 5dB &\footnotesize -5dB &\footnotesize 0dB &\footnotesize 5dB &\footnotesize -5dB &\footnotesize 0dB &\footnotesize 5dB \\
\hline 
Gerkmann        &0.61  &0.56  &0.49   &0.57  &0.53   &0.51  &0.55  &0.52  &0.46   \\
EM-ML           &0.51  &0.43  &0.41   &0.58  &0.56   &0.52  &0.49  &0.47  &0.46   \\
FTW-RBMs        &0.56  &0.53  &0.50   &0.47  &0.45   &0.44  &0.51  &0.48  &0.45   \\
\scriptsize EFTW-RBMs             &\textbf{0.45} &\textbf{0.42}  &\textbf{0.39} &\textbf{0.44}  &\textbf{0.43} &\textbf{0.40} &\textbf{0.42} &\textbf{0.40} &\textbf{0.37} \\
\hline 
\end{tabular*}
}
\end{threeparttable}
\vspace{-3mm}
\label{snoise2} 
\end{table}
\end{center}

\section{Conclusion}
In this letter, we propose a EFTW-RBMs model for SD. This gated RBMs approach can effectively introduce the frame-wise correlation and retains long-term speech features. By applying weight coefficients and a threshold function, the enhanced input units help to reconstruct the speech components  more robust, which significantly improves the SPP estimation in the T-F domain. The implementation of the proposed model reduces parameters and concentrates the energy of speech features by using nonnegative regularization. The evaluation results show that EFTW-RBMs demonstrate advantages over other state-of-the-arts in both 1D SD and 2D SD cases. The future work will focus on improving the selection of gated features, extending current enhanced FRBM into a stacked deep model, and promoting current model into a dynamically updated online model.
\ifCLASSOPTIONcaptionsoff
  \newpage
\fi

\bibliographystyle{IEEEtran}
\bibliography{references}








\end{document}